\input harvmac

\def\K3{{\bf K3}}
\def\journal#1&#2(#3){\unskip, \sl #1\ \bf #2 \rm(19#3) }
\def\andjournal#1&#2(#3){\sl #1~\bf #2 \rm (19#3) }

\def\bar{\overline}

\def\tilde{\widetilde}

\def\frac#1#2{{#1\over#2}}

\def\inbar{\,\vrule height1.5ex width.4pt depth0pt}
\def\IC{\relax\hbox{$\inbar\kern-.3em{\rm C}$}}
\def\IR{\relax{\rm I\kern-.18em R}}
\def\IP{\relax{\rm I\kern-.18em P}}

%
%

%
\catcode`\@=11
\def\slash#1{\mathord{\mathpalette\c@ncel{#1}}}
\overfullrule=0pt

\def\underrel#1\over#2{\mathrel{\mathop{\kern\z@#1}\limits_{#2}}}

\catcode`\@=12


%

\def\unit{\relax{\rm 1\kern-.26em I}}
\def\nada{\relax{\rm 0\kern-.30em l}}
\def\tilde{\widetilde}

\def\alphadot{{\dot \alpha}}



\noblackbox
\def\IL{\relax{\rm I\kern-.18em L}}
\def\IH{\relax{\rm I\kern-.18em H}}
\def\IR{\relax{\rm I\kern-.18em R}}
\def\IC{\relax\hbox{$\inbar\kern-.3em{\rm C}$}}
\def\IZ{\relax\ifmmode\mathchoice
{\hbox{\cmss Z\kern-.4em Z}}{\hbox{\cmss Z\kern-.4em Z}} {\lower.9pt\hbox{\cmsss Z\kern-.4em Z}}
{\lower1.2pt\hbox{\cmsss Z\kern-.4em Z}}\else{\cmss Z\kern-.4em Z}\fi}

\def\CJ {{\cal J}}

\def\CL {{\cal L}}


\font\manual=manfnt \def\dbend{\lower3.5pt\hbox{\manual\char127}}

\def\IZ{\relax\ifmmode\mathchoice
{\hbox{\cmss Z\kern-.4em Z}}{\hbox{\cmss Z\kern-.4em Z}} {\lower.9pt\hbox{\cmsss Z\kern-.4em Z}}
{\lower1.2pt\hbox{\cmsss Z\kern-.4em Z}}\else{\cmss Z\kern-.4em Z}\fi}

\def\alphadot{{\dot \alpha}}

\def\ibar{{\bar i}}

\def\bar{\overline}

\def\pa{\partial}

\def\rt2{\sqrt{2}}
\def\irt2{{1\over\sqrt{2}}}

\def\slashchar#1{\setbox0=\hbox{$#1$}           
   \dimen0=\wd0                                 
   \setbox1=\hbox{/} \dimen1=\wd1               
   \ifdim\dimen0>\dimen1                        
      \rlap{\hbox to \dimen0{\hfil/\hfil}}      
      #1                                        
   \else                                        
      \rlap{\hbox to \dimen1{\hfil$#1$\hfil}}   
      /                                         
   \fi}

\def\foursqr#1#2{{\vcenter{\vbox{
    \hrule height.#2pt
    \hbox{\vrule width.#2pt height#1pt \kern#1pt
    \vrule width.#2pt}
    \hrule height.#2pt
    \hrule height.#2pt
    \hbox{\vrule width.#2pt height#1pt \kern#1pt
    \vrule width.#2pt}
    \hrule height.#2pt
        \hrule height.#2pt
    \hbox{\vrule width.#2pt height#1pt \kern#1pt
    \vrule width.#2pt}
    \hrule height.#2pt
        \hrule height.#2pt
    \hbox{\vrule width.#2pt height#1pt \kern#1pt
    \vrule width.#2pt}
    \hrule height.#2pt}}}}
\def\psqr#1#2{{\vcenter{\vbox{\hrule height.#2pt
    \hbox{\vrule width.#2pt height#1pt \kern#1pt
    \vrule width.#2pt}
    \hrule height.#2pt \hrule height.#2pt
    \hbox{\vrule width.#2pt height#1pt \kern#1pt
    \vrule width.#2pt}
    \hrule height.#2pt}}}}
\def\sqr#1#2{{\vcenter{\vbox{\hrule height.#2pt
    \hbox{\vrule width.#2pt height#1pt \kern#1pt
    \vrule width.#2pt}
    \hrule height.#2pt}}}}

\def\figin{\epsfcheck\figin}\def\figins{\epsfcheck\figins}
\def\epsfcheck{\ifx\epsfbox\UnDeFiNeD
\message{(NO epsf.tex, FIGURES WILL BE IGNORED)}
\gdef\figin##1{\vskip2in}\gdef\figins##1{\hskip.5in}
\else\message{(FIGURES WILL BE INCLUDED)}%
\gdef\figin##1{##1}\gdef\figins##1{##1}\fi}
\def\DefWarn#1{}
\def\figinsert{\goodbreak\midinsert}
\def\ifig#1#2#3{\DefWarn#1\xdef#1{fig.~\the\figno}
\writedef{#1\leftbracket fig.\noexpand~\the\figno}%
\figinsert\figin{\centerline{#3}}\medskip\centerline{\vbox{\baselineskip12pt \advance\hsize by
-1truein\noindent\footnotefont{\bf Fig.~\the\figno:\ } \it#2}}
\bigskip\endinsert\global\advance\figno by1}


\def\ibar{{\bar i}}

\lref\GatesNR{
  S.~J.~Gates, M.~T.~Grisaru, M.~Rocek and W.~Siegel,
  ``Superspace, or one thousand and one lessons in supersymmetry,''
  Front.\ Phys.\  {\bf 58}, 1 (1983)
  [arXiv:hep-th/0108200].
}

\lref\KalloshVE{
  R.~Kallosh, L.~Kofman, A.~D.~Linde and A.~Van Proeyen,
  ``Superconformal symmetry, supergravity and cosmology,''
  Class.\ Quant.\ Grav.\  {\bf 17}, 4269 (2000)
  [Erratum-ibid.\  {\bf 21}, 5017 (2004)]
  [arXiv:hep-th/0006179].
}

\lref\DvaliZH{
  G.~Dvali, R.~Kallosh and A.~Van Proeyen,
  ``D-term strings,''
  JHEP {\bf 0401}, 035 (2004)
  [arXiv:hep-th/0312005].
}

\lref\FischlerZK{
  W.~Fischler, H.~P.~Nilles, J.~Polchinski, S.~Raby and L.~Susskind,
  ``Vanishing Renormalization Of The D Term In Supersymmetric U(1) Theories,''
  Phys.\ Rev.\ Lett.\  {\bf 47}, 757 (1981).
}

\lref\WittenNF{
  E.~Witten,
  ``Dynamical Breaking Of Supersymmetry,''
  Nucl.\ Phys.\  B {\bf 188}, 513 (1981).
}

\lref\ElvangJK{
  H.~Elvang, D.~Z.~Freedman and B.~Kors,
  ``Anomaly cancellation in supergravity with Fayet-Iliopoulos couplings,''
  JHEP {\bf 0611}, 068 (2006)
  [arXiv:hep-th/0606012].
}

\lref\ShifmanZI{
  M.~A.~Shifman and A.~I.~Vainshtein,
  ``Solution of the Anomaly Puzzle in SUSY Gauge Theories and the Wilson
  Operator Expansion,''
  Nucl.\ Phys.\  B {\bf 277}, 456 (1986)
  [Sov.\ Phys.\ JETP {\bf 64}, 428 (1986\ ZETFA,91,723-744.1986)].
}

\lref\BarbieriAC{
  R.~Barbieri, S.~Ferrara, D.~V.~Nanopoulos and K.~S.~Stelle,
  ``Supergravity, R Invariance And Spontaneous Supersymmetry Breaking,''
  Phys.\ Lett.\  B {\bf 113}, 219 (1982).
}

\lref\SeibergVC{
  N.~Seiberg,
  ``Naturalness Versus Supersymmetric Non-renormalization Theorems,''
  Phys.\ Lett.\  B {\bf 318}, 469 (1993)
  [arXiv:hep-ph/9309335].
}

\lref\DineXK{
  M.~Dine, N.~Seiberg and E.~Witten,
  ``Fayet-Iliopoulos Terms in String Theory,''
  Nucl.\ Phys.\  B {\bf 289}, 589 (1987).
}

\lref\FreedmanUK{
  D.~Z.~Freedman,
  ``Supergravity With Axial Gauge Invariance,''
  Phys.\ Rev.\  D {\bf 15}, 1173 (1977).
}

\lref\DasPU{
  A.~Das, M.~Fischler and M.~Rocek,
  ``Superhiggs Effect In A New Class Of Scalar Models And A Model Of Super
  QED,''
  Phys.\ Rev.\  D {\bf 16}, 3427 (1977).
}

      \lref\deWitWW{
        B.~de Wit and P.~van Nieuwenhuizen,
       ``The Auxiliary Field Structure In Chirally Extended Supergravity,''
        Nucl.\ Phys.\  B {\bf 139}, 216 (1978).
      }

\lref\BinetruyHH{
  P.~Binetruy, G.~Dvali, R.~Kallosh and A.~Van Proeyen,
  ``Fayet-Iliopoulos terms in supergravity and cosmology,''
  Class.\ Quant.\ Grav.\  {\bf 21}, 3137 (2004)
  [arXiv:hep-th/0402046].
}

\lref\FerraraDH{
  S.~Ferrara, L.~Girardello, T.~Kugo and A.~Van Proeyen,
  ``Relation Between Different Auxiliary Field Formulations Of N=1 Supergravity
  Coupled To Matter,''
  Nucl.\ Phys.\  B {\bf 223}, 191 (1983).
}

\lref\WeinbergUV{
  S.~Weinberg,
  ``Non-renormalization theorems in non-renormalizable theories,''
  Phys.\ Rev.\ Lett.\  {\bf 80}, 3702 (1998)
  [arXiv:hep-th/9803099].
}

\lref\FerraraPZ{
  S.~Ferrara and B.~Zumino,
  ``Transformation Properties Of The Supercurrent,''
  Nucl.\ Phys.\  B {\bf 87}, 207 (1975).
}

\lref\IntriligatorCP{
  K.~A.~Intriligator and N.~Seiberg,
  ``Lectures on Supersymmetry Breaking,''
  Class.\ Quant.\ Grav.\  {\bf 24}, S741 (2007)
  [arXiv:hep-ph/0702069].
}

\lref\DineTA{
  M.~Dine,
  ``Fields, Strings and Duality: TASI 96,''
 eds. C.~Efthimiou and B. Greene (World Scientific, Singapore, 1997).
   }

\lref\FerraraPZ{
  S.~Ferrara and B.~Zumino,
  ``Transformation Properties Of The Supercurrent,''
  Nucl.\ Phys.\  B {\bf 87}, 207 (1975).
}

\lref\WittenBZ{
  E.~Witten,
  ``New Issues In Manifolds Of SU(3) Holonomy,''
  Nucl.\ Phys.\  B {\bf 268}, 79 (1986).
}

\lref\SeibergVC{
  N.~Seiberg,
 ``Naturalness Versus Supersymmetric Non-renormalization Theorems,''
  Phys.\ Lett.\  B {\bf 318}, 469 (1993)
  [arXiv:hep-ph/9309335].
}

\lref\ORaifeartaighPR{
  L.~O'Raifeartaigh,
  ``Spontaneous Symmetry Breaking For Chiral Scalar Superfields,''
  Nucl.\ Phys.\  B {\bf 96}, 331 (1975).
}

\lref\FayetJB{
  P.~Fayet and J.~Iliopoulos,
  ``Spontaneously Broken Supergauge Symmetries and Goldstone Spinors,''
  Phys.\ Lett.\  B {\bf 51}, 461 (1974).
}

\lref\GreenSG{
  M.~B.~Green and J.~H.~Schwarz,
  ``Anomaly Cancellation In Supersymmetric D=10 Gauge Theory And Superstring
  Theory,''
  Phys.\ Lett.\  B {\bf 149}, 117 (1984).
}

\lref\ChamseddineGB{
  A.~H.~Chamseddine and H.~K.~Dreiner,
  ``Anomaly Free Gauged R Symmetry In Local Supersymmetry,''
  Nucl.\ Phys.\  B {\bf 458}, 65 (1996)
  [arXiv:hep-ph/9504337].
}

\lref\CastanoCI{
  D.~J.~Castano, D.~Z.~Freedman and C.~Manuel,
  ``Consequences of supergravity with gauged U(1)-R symmetry,''
  Nucl.\ Phys.\  B {\bf 461}, 50 (1996)
  [arXiv:hep-ph/9507397].
}

\lref\WittenHU{
  E.~Witten and J.~Bagger,
  ``Quantization Of Newton's Constant In Certain Supergravity Theories,''
  Phys.\ Lett.\  B {\bf 115}, 202 (1982).
}

\lref\BaggerFN{
  J.~Bagger and E.~Witten,
  ``The Gauge Invariant Supersymmetric Nonlinear Sigma Model,''
  Phys.\ Lett.\  B {\bf 118}, 103 (1982).
}


\Title{
} {\vbox{\centerline{Comments on the Fayet-Iliopoulos Term}
\centerline{}
\centerline {in Field Theory and Supergravity}}}
\medskip

\centerline{\it Zohar Komargodski and Nathan Seiberg}
\bigskip
\centerline{School of Natural Sciences}
\centerline{Institute for Advanced Study}
\centerline{Einstein Drive, Princeton, NJ 08540}

\smallskip

\vglue .3cm

\bigskip
\noindent
A careful analysis of the Fayet-Iliopoulos (FI) model shows that its energy momentum tensor and supersymmetry current are not gauge invariant.  Since the corresponding charges are gauge invariant, the model is consistent.  However, our observation about the currents gives a new perspective on its restrictive renormalization group flow and explains why FI-terms never appear in dynamical supersymmetry breaking.  This lack of gauge invariance is at the root of the complications of coupling the model to supergravity.  We show that this is possible only if the full supergravity theory (including all higher derivative corrections) has an additional exact continuous global symmetry.  A consistent quantum gravity theory cannot have such symmetries and hence FI-terms cannot appear.  Our results have consequences for various models of particle physics and cosmology.

\Date{April 2009}

\newsec{Introduction and Summary}

Supersymmetry (SUSY) breaking arises in two somewhat different forms. One is F-term breaking whose prime example is the O'Raifeartaigh model \ORaifeartaighPR\ and the other is D-term breaking whose prime example is the FI model \FayetJB.  Models in the latter class are based on a gauge symmetry which we will denote by $U(1)_{FI}$.

The energy momentum tensor $T_{\mu \nu}$ and the supersymmetry current $S_{\mu\alpha}$ reside in a supersymmetry multiplet which was first constructed in \FerraraPZ.\foot{We do not discuss here alternative multiplets, such as those in section 7 of \GatesNR.}  Studying this multiplet, we find that the usual FI-term renders this multiplet not gauge invariant.  More precisely, a supersymmetric gauge theory has a large gauge symmetry group.  It is common to fix Wess-Zumino (WZ) gauge in which the remaining gauge freedom is ordinary gauge transformations.  $T_{\mu \nu}$ and $S_{\mu\alpha}$ are gauge invariant under this remaining gauge freedom but not under the full gauge symmetry of the theory.
This peculiarity can be traced to the fact that the Lagrangian in superspace is not gauge invariant.

It is worth emphasizing that some models have field dependent FI-terms \DineXK. These look approximately like the ordinary FI-term for some purposes.  However, they differ from the original ``field independent" FI-terms in two crucial ways.  First, they inevitably lead to the spontaneous breaking of the gauged $U(1)_{FI}$ symmetry and hence its massive gauge multiplet can be integrated out. Second, the corresponding Lagrangian is gauge invariant in superspace, a property which does not hold in the presence of field independent FI-terms.  In fact, the name ``field dependent FI-terms'' is misleading -- such terms should not be called FI-terms.\foot{We thank E.~Witten for a useful comment on this point.} Our comments in this note will mostly apply to the genuine (field independent) FI-terms.

Our observation about the lack of gauge invariance of the supersymmetry current for the usual FI-terms will allow us to re-derive the non-renormalization of FI-terms \refs{\FischlerZK\ShifmanZI\DineTA-\WeinbergUV}.  The authors of \FischlerZK\ gave a perturbative diagrammatic proof.  Reference \ShifmanZI\ used holomorphy, and more generally, \refs{\DineTA,\WeinbergUV} followed the non-renormalization theorem of \SeibergVC.  Similarly, if the short distance theory does not have an FI-term, the same is true at all length scales. Furthermore, our analysis shows that if a $U(1)$ gauge field emerges at low energies from the dynamics, it cannot appear with an FI-term.  This explains why all calculable models of dynamical supersymmetry breaking exhibit F-term breaking rather than D-term breaking.  (For a recent review and earlier references see e.g.~\IntriligatorCP.)

Superficially, if the currents $T_{\mu \nu}$ and $S_{\mu\alpha}$ are not $U(1)_{FI}$ gauge invariant, one cannot gauge them; i.e.\ one cannot couple the model to supergravity.  Nevertheless \refs{\FreedmanUK\DasPU-\deWitWW} succeeded to couple the FI model to supergravity.  This construction was further explored in \refs{\BarbieriAC,\FerraraDH} and it was emphasize that the $U(1)_{FI}$ charges of the various fields are shifted by an amount proportional to $\xi \over M_P^2$.

This shift of the charges has raised some doubts about the validity of the general construction.  The authors of \refs{\BarbieriAC,\FerraraDH}  pointed out that the generic rigid supersymmetry theory cannot be coupled to gravity unless it has an exact global $U(1)_R$ symmetry.  We will argue that the need for this symmetry and the corresponding shift of the charges directly follows from the lack of $U(1)_{FI}$ gauge invariance of $T_{\mu \nu}$ and $S_{\mu\alpha}$.  Also, Witten \WittenBZ\ argued that if the theory has magnetic monopoles, the shift of the charges is incompatible with Dirac quantization and hence it is inconsistent.  Finally, \refs{\ChamseddineGB\CastanoCI\BinetruyHH-\ElvangJK} showed that imposing the absence of anomalies severely constrains the model with its shifted charges. These constraints on supergravity theories with FI-terms are strong but not sufficient to rule out this possibility.

Below we will show that the additional $U(1)_R$ global symmetry must be present not only in the rigid limit, but it must be an exact symmetry of the full quantum gravity theory.  This fact is in contradiction with general rules about lack of global continuous symmetries in a consistent theory of gravity.  Therefore, we conclude that such FI-terms cannot arise in supergravity.

Our observations thus explain why, despite many efforts, nobody could come up with a string construction containing an FI-term in the low energy limit.

Throughout our analysis we assume that the FI-term $\xi $ is parametrically smaller than $M_P^2$.  If $\xi \sim M_P^2$ a description in terms of a supergravity theory with a $U(1)_{FI}$ gauge symmetry cannot be complete because it includes Planck-scale physics.

\newsec{The FI-Term in Supersymmetric Field Theories}

\subsec{The Supercurrent Multiplet}

In supersymmetric field theories the supercurrent $S_{\mu\alpha}$ and the energy-momentum tensor $T_{\mu\nu}$ reside in the same multiplet \FerraraPZ.  Here we review this construction emphasizing the facts most relevant to our purposes.

This multiplet is a real superfield which is a Lorentz vector $\CJ_{\alpha\dot\alpha}$.\foot{Our convention is that for every vector $\ell_\mu$,
$$\ell_{\alpha\dot\alpha}=-2\sigma^\mu_{\alpha\dot\alpha}\ell_\mu,\qquad \ell_\mu={1\over 4}\bar\sigma_\mu^{\dot\alpha\alpha}\ell_{\alpha\dot\alpha}~.$$}  The conservation equation is \FerraraPZ
\eqn\defeq{\bar D^{\dot\alpha}\CJ_{\alpha\alphadot}=D_\alpha X~,}
with $X$ some chiral superfield.  As we said above, we do not discuss alternative multiplets such as those of \GatesNR.  

The solution of this equation in components is
\eqn\currentsuperfield{\eqalign{
\CJ_\mu=& j_\mu+\theta^\alpha\left(S_{\mu\alpha}+{1\over 3}(\sigma_\mu\bar\sigma^\rho S_\rho)_\alpha\right)+\bar\theta_{\dot\alpha}
\left(\bar S_\mu^{\dot\alpha}+{1\over 3}\epsilon^{\dot\alpha\dot\beta}(\bar S_\rho\bar\sigma^\rho\sigma_\mu)_{\dot\beta} \right)\cr
&+(\theta\sigma^\nu\bar\theta)\left(2T_{\nu\mu}-{2\over 3}\eta_{\nu\mu}T-{1\over4}\epsilon_{\nu\mu\rho\sigma}\pa^{[\rho}j
^{\sigma]}\right)+\cdots~,\cr
\  \   \ \  \pa^\mu T_{\mu\nu} & =\pa^\mu S_{\mu\alpha}=0~,}}
where the ellipses stand for terms which are unimportant to our discussion.  $j_\mu$, $S_{\mu \alpha}$ and $T_{\mu\nu}$ are an R-symmetry current, the supercurrent and the energy-momentum tensor.

In fact, equation \defeq\ does not uniquely determine the solution. If $\CJ_{\alpha\dot\alpha}$ and $X$ satisfy~\defeq, then
\eqn\newpair{\eqalign{
&\CJ_{\alpha\dot\alpha}'= \CJ_{\alpha\dot\alpha}+i\pa_{\alpha\dot\alpha}(Y-\bar Y) \cr
& X'= X-{1\over 2}\bar D^2\bar Y}}
also satisfy \defeq\ for any chiral superfield $Y$.  This ambiguity is familiar from nonsupersymmetric theories.  In components the ambiguity \newpair\ corresponds to changing the improvement terms
\eqn\enemo{\eqalign{
&S'_{\mu \alpha} = S_{\mu \alpha} + 2i \left(\sigma_{\mu \nu}\right)_\alpha^\beta  \partial^\nu Y\bigr|_{\theta^\beta} \cr
& T'_{\mu\nu}=T_{\mu\nu}-\left(\pa_\mu\pa_\nu-\eta_{\mu\nu}\pa^2\right){\rm Re} Y\bigr| ~,}}
where $Y=Y\bigr|+\theta^\beta Y\bigr|_{\theta^\beta}+\cdots~$.  These do not change the supersymmetry charge $Q_\alpha$ and the four-momentum $P_\mu$. (Because of the possible vacuum energy density, the energy $P_0$ could diverge.  Then this comment applies to the finite volume system.)

As an example, consider the general sigma model
\eqn\sigmamodel{\int d^4\theta K(\Phi _i, \bar\Phi_\ibar )
 + \int d^2 \theta
 W(\Phi_i) + \int d^2 \bar \theta \ \bar W(\bar \Phi_\ibar)~,}
where $\Phi_i$ are chiral superfields.
The supercurrent is
\eqn\supercurrensm{\CJ_{\alpha \alphadot}= 2g^{i \ibar}( D_\alpha  \Phi_i)(\bar D_\alphadot \bar \Phi_\ibar)
 - {2\over 3}[D_\alpha, \bar D_\alphadot ] K~.}
The physics is not affected by K\"ahler transformation
\eqn\kahlertrans{K'(\Phi _i, \bar\Phi_\ibar )=K(\Phi _i, \bar\Phi_\ibar )+F(\Phi_i)+\bar F(\bar \Phi_\ibar)~,}
but the superfield $\CJ_{\alpha\alphadot}$ in \supercurrensm\ transforms as
\eqn\supctransf{\CJ'_{\alpha \alphadot}=\CJ_{\alpha \alphadot}+i{2\over 3}\pa_{\alpha\alphadot}\left(F(\Phi_i)-\bar F(\bar\Phi_\ibar)\right)~.}
We recognize this change as the ambiguity pointed out in \newpair.
Hence, we see that the freedom in choosing a solution to \defeq\ is related to K\"ahler transformations.

\subsec{The Case of an FI D-Term.}

Let us analyze the superfield $\CJ_{\alpha\alphadot}$ in the presence of an FI-term. We begin by considering a free theory of a single vector multiplet and an explicit FI-term
\eqn\supFI{W={1\over 4g^2}W_\alpha^2,\qquad K=\xi V~.}
The supercurrent is
\eqn\supercurrentfreeD{\CJ_{\alpha\dot\alpha}= -{4\over g^2}W_\alpha \bar W_{\dot\alpha}-{2\over 3}\xi [D_\alpha,\bar D_\alphadot]V~.}
Indeed,  \defeq\ can be shown to be satisfied with
\eqn\XfreeD{X=-{\xi\over 3}\bar D^2V~.}

In the presence of an FI-term, general gauge transformations \eqn\susyg{V'=V+i\left(\Lambda-\bar\Lambda\right)}
induce K\"ahler transformations. Therefore, in light of \supctransf\ it is not surprising that the supercurrent superfield \supercurrentfreeD\ is not gauge invariant. It is also clear that this is a general property of the FI-term and not just a particular feature of the free theory. Unlike the harmless ambiguity associated with K\"ahler transformations \supctransf, the lack of gauge invariance has profound physical consequences, as we will soon see.

The supercurrent and the energy momentum tensor are now not gauge invariant under general gauge transformations \susyg:
\eqn\enemoa{\eqalign{
&S'_{\mu \alpha} = S_{\mu \alpha} - {4\xi\over 3} \left(\sigma_{\mu \nu}\right)_\alpha^\beta  \partial^\nu \Lambda\bigr|_{\theta^\beta}\cr
&T'_{\mu\nu}=T_{\mu\nu}+{2\xi\over 3}\left(\pa_\mu\pa_\nu-\eta_{\mu\nu}\pa^2\right){\rm Im} \Lambda\bigr|  ~,}}
where $\Lambda\bigr|$ and $\Lambda\bigr|_{\theta ^\beta} $ are the bottom and the $\theta^\beta$ components of the superfield $\Lambda$.
Note that as we explained in \enemo, the supersymmetry charge $Q_\alpha$ and the four-momentum $P_\mu$ are gauge invariant.  Therefore, this lack of gauge invariance does not mean that the theory is inconsistent.

Let us consider the system in Wess-Zumino gauge.  Here, the remaining gauge symmetry is ordinary gauge transformations with the parameter ${\rm Re} \Lambda\bigr|$.  It is clear from \enemoa\ that $T_{\mu\nu}$ and $S_{\mu \alpha}$ are invariant under this restricted set of transformations.  However, since this gauge choice breaks supersymmetry, all the conclusions we will derive below which stem from the fact that $\CJ_{\alpha\alphadot}$ is not gauge invariant are still valid.  One way to see that is to examine the R-current which is related to $T_{\mu\nu}$ and $S_{\mu \alpha}$ by supersymmetry transformations.  This current is the bottom component of \supercurrentfreeD
\eqn\bottomcompi{j_\mu={1\over g^2}\bar\lambda\bar\sigma_\mu\lambda-{2\over3}\xi A_\mu+\cdots~}
where the ellipses represent terms that vanish in the WZ gauge.  Clearly, $j_\mu$ is not gauge invariant even in the WZ gauge.

We will see that this lack of gauge invariance of the supercurrent multiplet provides a strong handle on the behavior of supersymmetric field theories with an FI-term and is especially useful in coupling these theories to supergravity.

\subsec{Consequences for Supersymmetric Field Theories}

Suppose a supersymmetric field theory has no FI-terms at high energy, where the theory is defined. Then, since the supercurrent exists and is gauge invariant, it must remain such throughout the whole renormalization group flow. Therefore, no FI-term can be present in the effective action at lower energies.  Similarly, no FI-term can be generated at any order in perturbation theory and even not due to non-perturbative effects.\foot{A well known exception is the anomalous situation when the sum of the $U(1)_{FI}$ charges does not vanish and a quadratically divergent FI-term is generated at one loop \refs{\WittenNF,\FischlerZK}.} If the theory flows through a regime with strong dynamics and one has a dual description, it might be that there is an emergent $U(1)$ gauge symmetry at low energies. However, gauge invariance prevents this gauge multiplet from having a nonzero FI-term.

This explains why all the known examples of dynamical SUSY breaking  are always predominantly F-term driven. Further, if there is an FI-term at the tree-level, its value is not renormalized along the renormalization group flow, since the non-gauge invariance of the supercurrent is preserved.  (One way to study it is in terms of a line integral of the R-current which ends on a source.)

These are the main points in our field theory discussion.  The rest of this section presents various examples and comments.

If the theory contains a charged matter field $\Phi$ which Higgses $U(1)_{FI}$, then one might try to correct the FI-term by writing
\eqn\correcting{K'=\xi \left(V+c\ln(|\Phi|^2)\right)~,}
with an appropriate coefficient $c$.   This modification does not affect the theory because it is a (singular) K\"ahler transformation, but it renders the current superfield gauge invariant.  However, since this transformation is singular at $\Phi=0$ we have to be more careful.  If the point $\Phi=0$ is at finite distance (in the K\"ahler metric) from the vacuum, such a singularity is unacceptable.  If instead the point $\Phi=0$ is at infinite distance, a modification like \correcting\ solves the problem.  However, in this case the gauge symmetry $U(1)_{FI}$ is Higgsed in the entire field space and it is meaningless to refer to it as an FI-term.

A similar situation arises in string theory~\DineXK, where the the Green-Schwarz mechanism~\GreenSG\ leads to a field dependent FI-term
\eqn\stringD{ K \sim \ln\left(Y+\bar Y+V\right) ~.}
The fields $Y,\bar Y$ transform additively under gauge transformations (they are essentially axions) and hence \stringD\ is gauge invariant and so is the supercurrent multiplet.  The situation here is similar to \correcting\ as the gauge symmetry is always Higgsed (the singularity is at infinite distance).  More quantitatively, if supersymmetry is unbroken, the scalar in $Y$ is eaten in the Higgs mechanism and the mass of $Y$ is the same as that of $V$.  Therefore, there is no energy range where the field $Y$ is heavy and can be integrated out, while $V$ is light and has an FI-term.

Another way FI-terms can arise is from non-renormalizable terms such as
\eqn\MSSMD{K={1\over M^2}X\bar X D^\alpha W_\alpha~,}
where $X$ is a spurion, $X=\theta^2F$, and $W_\alpha$ is the field strength chiral superfield. This leads to an effective FI-term at low energies of the order $\xi\sim F^2/M^2$. However, clearly, the underlying theory is gauge invariant in superspace and possesses a well defined supercurrent. Indeed, \MSSMD\  can be removed by a change of variables of the form
\eqn\vchange{V'=V+{1\over M^2}X\bar X~,}
such that $V'$ has no D-term VEV. Therefore, this D-term is essentially ``fake," as all its effects can be seen  in a theory where there is no D-term (but the K\"ahler potential is different).  Clearly, in such circumstances the D-terms are comparable or smaller than some F-terms.  One well known example of this phenomenon appears in the $U(1)_Y$ hypercharge gauge symmetry of the MSSM.

\newsec{The FI-Term in Supergravity}

Coupling a rigid supersymmetric theory with a nonzero FI-term  $\xi$ to supergravity is nontrivial \refs{\FreedmanUK - \ElvangJK}.  Here we review this construction from our perspective emphasizing the fact that the supercurrent is not gauge invariant.

Let us first consider the simple case of $\xi=0$.  A constructive way of obtaining supergravity theories is to gauge the supercurrent multiplet $\CJ_{\alpha\alphadot}$.  The bottom component of this multiplet \currentsuperfield\ is the current $j_\mu$ of a $U(1)_R$ transformation under which all the chiral fields have $R=2/3$.  Gauging the supersymmetry current $\CJ_{\alpha\alphadot}$ includes coupling a gauge field $B_\mu$ to this $j_\mu$.  The fact that $j_\mu$ is not necessarily conserved is typically solved by introducing an additional chiral multiplet, a compensator $Y$, which carries charge $R=2/3$.  This renders the theory R-invariant and the corresponding current $j_\mu$  is conserved. (This construction is reviewed, for instance, in \KalloshVE.) The original theory is obtained by setting
\eqn\gaugefixing{Y=Y^*=\sqrt3 M_P~.}
This equation can be interpreted as a unitary gauge choice for the gauged $U(1)_R$ making $B_\mu$ heavy and decoupled.

In more detail, we consider the superpotential of the original theory,
$W(\Phi_i,m_j)$, where we have displayed the mass parameters that may appear in the superpotential explicitly.  Adding an additional chiral field $Y$ such a theory can be made R-symmetric. We write
 \eqn\zdef{Z_i={\Phi_i\over Y},\qquad \tilde m_j={m_j\over \sqrt3 M_P}~,}
and now $Z_i$  are dimensionless and neutral under $j_\mu$. We can thus modify the superpotential
 \eqn\conservedcurrent{W(\Phi_i,m_j)\rightarrow Y^3W(Z_i,\tilde m_j)~.}
This theory is R-symmetric and reduces to the original one upon substituting the VEV~\gaugefixing.

We now turn to the more interesting case of a SUSY field theory with a non-zero FI-term $\xi$ for a $U(1)_{FI}$ gauge symmetry with a gauge field $A_\mu$ . Now the current $j_\mu$ is not gauge invariant; as in \bottomcompi, it contains
\eqn\currentnong{j_\mu=\cdots-{2\over 3}\xi A_\mu~.}
For $\xi=0$ we corrected the non-conservation of $j_\mu$ by adding the compensator $Y$. With an FI-term the same compensator can fix the problem of gauge invariance.  (We do not have the freedom to introduce an additional  ``compensator'' for this purpose, because that would change the particle content of the theory.)  We assign $U(1)_{FI}$ charge the chiral multiplet $Y$ (chosen to agree with \BinetruyHH)
\eqn\transf{\delta_{U(1)_{FI}} Y=i{\xi\over 3M_P^2 }Y~.}
Roughly speaking, this makes the current gauge invariant because now the FI-term essentially arises from $3M_P^2\log|Y|^2 + \xi V $ in the K\"ahler potential.

However, the assignment \transf\ is meaningful only if it leaves the full theory invariant.  We see from the form \conservedcurrent\ that this indeed the case, if and only if the original rigid theory had an R-symmetry.\foot{This R-symmetry should not be confused with the R-symmetry coupled to $B_\mu$ we gauged above.} Similarly, it is clear that the K\"ahler potential should also have the same R-symmetry.  Denoting the R-charges by $R(\Phi_i)=R_i $, we must shift the $U(1)_{FI}$ charges as
          \eqn\charges{
 \matrix{
                         &  \widetilde{U(1)}_{FI} \cr
          & \cr
         \phi_i &   Q_i+{R_i\xi\over 2M_P^2} -{\xi\over 3M_P^2}\cr
         \psi_{\phi_i} &Q_i+{R_i\xi\over 2M_P^2} -{\xi\over 3M_P^2} \cr
         Y &    -{\xi\over 3M_P^2}  \cr
         \psi_Y &  -{\xi\over 3M_P^2}   \cr
          }}
Note that the shift of the charges is determined by $R_i$.  However, the symmetry $\widetilde{U(1)}_{FI}$  is an ordinary gauge symmetry which is not an R-symmetry.

At this stage our theory has a gauged $U(1)_R \times \widetilde{U(1)}_{FI}$ symmetry with the two gauge fields $B_\mu$ and $A_\mu$.
The VEV of $Y$ \gaugefixing\ Higgses this symmetry group to a subgroup $U(1)_{FI}^\xi$ with charges

\eqn\chargesi{
 \matrix{
                         &  U(1)_{FI}^\xi \cr
          & \cr
         \phi_i &  Q_i+R_i{\xi\over2M_P^2}  \cr
         \psi_{\phi_i} & Q_i+(R_i-1){\xi\over2M_P^2}   \cr
         \lambda &  {\xi\over 2M_P^2} \cr
         \psi_\mu & {\xi\over 2M_P^2} \cr
          }}
where $\lambda $ and $\psi_\mu$ are gauginos and the gravitino. We have not displayed the charges of the fields in the $Y$ multiplet as they are not part of the low energy description.  Note that $ U(1)_{FI}^\xi$ is a gauged R-symmetry under which the supercharge, the superspace coordinate $\theta$ and the gravitino are charged.

As we have mentioned in the introduction, this shift of the charges looks problematic for many reasons including charge quantization \WittenBZ\ and anomalies \refs{\ChamseddineGB\CastanoCI\BinetruyHH-\ElvangJK} (see also references therein). In addition, only theories with an exact R-symmetry in the rigid limit can be coupled to supergravity \BarbieriAC. These are strong constraints but leave room for models that satisfy all of them. There is nothing wrong with supersymmetric field theories that have an exact R-symmetry and by properly choosing the field content also the anomalies can be canceled.

It is straightforward to extend the previous analysis to the full theory including its higher derivative corrections.  We assume that such a theory can be written in terms of the superfields of the rigid theory, the compensator $Y$ and the supergravity multiplet.  For $\xi=0$ we find a theory which has a gauge symmetry $U(1)_{FI}$.  We introduce the compensator $Y$ such that all terms are invariant under $U(1)_R$ with the scalars carrying $R=2/3$.  Then, for nonzero $\xi$ we need to assign $U(1)_{FI} $ transformation laws to $Y$.  In order to make the theory invariant we must have an exact {\it global} $U(1)_R$ symmetry with $R(\phi_i)=R_i$ and $R(Y)=0$. Then we can shift the $U(1)_{FI}$ charges as in \charges. Now we have an ordinary gauge symmetry (not an R-symmetry) $\widetilde{U(1)}_{FI}$ with a photon $A_\mu$, a local R-symmetry with gauge field $B_\mu$.  In addition we have an exact global symmetry which can be taken to be an R-symmetry or simply the original $U(1)_{FI}$ which does not act on $Y$.  Now we can gauge fix $Y$ and integrate out $B_\mu$.

An alternative way to say that is as follows.  The full theory has a Lagrangian ${\cal L}_\xi$ and a gauge symmetry $U(1)_{FI}^\xi$ generated by $\delta_\xi$.  Clearly,
\eqn\xiinva{\delta_\xi {\cal L}_\xi =0~.}
These transformations act linearly on the fields.  Let us keep $M_P$ fixed and expand in $\xi$ (keeping terms with all numbers of derivatives)
\eqn\xiexpansion{{\cal L}_\xi=\CL_0+\xi\CL_1+\cdots,\qquad \delta_\xi=\delta_0+\xi\delta_1+\cdots~.}
Since $\delta_\xi$ acts linearly on the fields for all $\xi$,
\eqn\commutrel{[\delta_1,\delta_0]=0~.}
The first two orders in the expansion of equation \xiinva\
\eqn\firstorder{\eqalign{
&\delta_0 {\cal L}_0=0\cr
& \delta_1\CL_0+\delta_0\CL_1=0~.}}
Acting on the second expression with $\delta_0$ and using \commutrel\ and \firstorder\ we find $\delta_0^2\CL_1=0$. Since this transformation is just a rotation of the fields (we can focus on space-time independent transformations for this argument) we conclude that
\eqn\conclusion{\delta_0\CL_1=\delta_1\CL_0=0~,}
namely, there is another global symmetry for $\xi=0$.  In fact, continuing this way to all orders in $\xi$ it is straightforward to see that the full theory with nonzero $\xi$ has $\delta_0$ as an exact global symmetry.

The conclusion that the full theory with the Lagrangian $\CL_\xi$ must have an exact global symmetry is problematic.  It violates general rules about the consistency of quantum gravity.  We conclude that this supergravity theory must be inconsistent and all supergravity theories must have vanishing FI-terms.
This explains why the efforts to find string theory models with FI-terms at low energies have been futile.

This understanding leads to a new derivation of some of the results in section 2.  In particular, the fact that FI-terms cannot be dynamically generated is obvious.  If the original theory can be coupled to supergravity, so should be its effective low energy theory.  Since we cannot have nonzero $\xi$ in supergravity, such a term cannot be generated.

\bigskip
\centerline{\bf Acknowledgements}
We would like to thank O.~Aharony, N.~Arkani-Hamed, T.~Banks, M.~Dine, H.~Elvang, D.~Freedman, J.~Maldacena, P.~Meade, M.~Rocek, D.~Shih, W.~Siegel, S.~Thomas, and E.~Witten for useful discussions. The work of ZK was supported in part by NSF grant PHY-0503584 and that of NS was supported in part by DOE grant DE-FG02-90ER40542. Any opinions, findings, and conclusions or recommendations expressed in this material are those of the author(s) and do not necessarily reflect the views of the funding agencies.

\listrefs
\end